# An Integrated Search Framework for Leveraging the Knowledge-Based Web Ecosystem

**Dengya Zhu**
Curtin Business School
Curtin University

**Shastri Lakshman Nimmagadda**
Curtin Business School
Curtin University
shastri.nimmagadda@curtin.edu.au

**Torsten Reiners**
Curtin Business School
Curtin University

**Amit Rudra**
Curtin Business School
Curtin University

## Abstract

The explosion of information constrains the judgement of search terms associated with Knowledge-Based Web Ecosystem (KBWE), making the retrieval of relevant information and its knowledge management challenging. The existing information retrieval (IR) tools and their fusion in a framework need attention, in which search results can effectively be managed. In this article, we demonstrate the effective use of information retrieval services by a variety of users and agents in various KBWE scenarios. An innovative Integrated Search Framework (ISF) is proposed, which utilises crawling strategies, web search technologies and traditional database search methods. Besides, ISF offers comprehensive, dynamic, personalized, and organization-oriented information retrieval services, ranging from the Internet, extranet, intranet, to personal desktop. In this empirical research, experiments are carried out demonstrating the improvements in the search process, as discerned in the conceptual ISF. The experimental results show improved precision compared with other popular search engines.

## 1 Introduction

Diverse domains and contexts make systems complex, effecting retrieval of data and information more challenging. The digital ecosystem (Gartner, 2017) is an open community which holds data from associated communities and at the same time produces consumable data to benefit others. More specifically, people and enterprises in a digital ecosystem share standardized digital content in various platforms, mutually beneficial to each other. Moore et al. (2018) assess the digital value in a case study with more than 3000 executives as "digital ecosystems are transforming the way their organisations deliver value" (Moore et al, page 5). However, data in digital ecosystems are distributive, complex, heterogeneous, and multidimensional (Barrows and Traverso, 2006). In other words, they have all the features of big data. For example, the volume of the data generated per hour in a digital ecosystem ranges from megabytes to gigabytes to terabytes; tens of thousands of bytes of data are transported per seconds, demonstrating velocity of data. For a variety of data, formats vary from, for example, emails to instant messages and images to streaming data.

The heterogeneity and multidimensionality of ecosystems and their data sources force the management information systems to make amendments, for creating scopes of better information retrieval methods and search functions, adaptable to a new information-access era. Information retrieval needs more specific search tools and formulations in addition to the presentation of search results in a way they can be better interpreted and analysed. Multiple information scenarios of web ecosystems make businesses shifting their focus to new flexible, re-configurable, and collaborative search models. Consequently, they should adapt to digitalization trends, and strive to leverage their data for competitive advantage. With the vast amount of data in hand, one of the distinct steps for organizations is to facilitate the data search process by easing the complexity of digital web ecosystems. For example, Commonwealth Bank of Australia (CBA) developed an app that assists users to search properties that they are interested in with sale price history as well as similar information of other properties nearby, and then with mortgage link points to CBA. After six months, consumers made more than 1.2 million property searches, and the app's return on investment is 109% (Weill and Woerner, 2015).

For exploring web-based digital ecosystems, relevant enterprise-level search tools/platforms are described in Barrows and Traverso (2006). The Apache Solar and Lucene are typical examples of the type of search tools, which provide services at an enterprise level (McCandless et al. 2010). They offer distributed indexing/searching services with high scalability, availability and extensibility features. The customized version may include an entity extractor, thesaurus, classification, filtering and other characteristics. *Desktop search* that focuses on retrieving local files residing on personal desktop computers, messages, emails, and browsing history, is another type of searching tool. The third type of such gadget is an *intranet search* engine, which crawls information from servers within an intranet to local networks. For resolving enterprise-scale problems, a search engine should support and combine tagging, categorization and navigation tools to improve the end-user experience. An enterprise metadata category – ontology-based metadata – can be built to define a metadata schema, to index a set of documents, and write a user interface for querying and displaying results. Even though automatic metadata extraction is never perfect, a user interface is needed to allow amendments and re-use of the metadata. In addition, an ecosystem that is supported by KBWE should satisfy scalability, security, metadata update, view privilege and query (search-term) optimization criteria (Albro, 2006).

The introduction has motivated us exploring the current literature to identify the research purpose in the contexts of information retrieval from not only large size data sources but also the type of support needed to retrieve the relevant facts and ascertain their evaluable precision. In addition, we review the existing IR designs with implementable framework, as described in the following sections.

## 2   Literature Review

Several researchers have described the existing concepts, tools, technologies and challenges to examine the data and IR from the World Wide Web (WWW). We examine the existing literature on managing large and complex datasets with traditional hardware and software systems and the frameworks that can process and retrieve required information with precision. The current information retrieval models and how they can replace the KBWE guided ISF and leverage their evaluations in knowledge management are described.

## 2.1　Information Retrieval from Big Data

The traditional database systems and analytic tools have weaknesses to process the Big Data, which is unstructured or semi-structured with a large amount of image and audio data (Haneef et al. 2018). As an example, managing heterogeneity of Big Data, in particular, data volumes and varieties with deliverable quality and precision through IR systems are challenging. For retrieving large size images, data related to audio and videos including social media posts and web blogs, tools for interrogation of big data, data warehouse technologies, and data analytics are all supportive in the information retrieval research. McCreadie et al. (2012) developed an information retrieval mechanism to improve the textual information using map/reduce to process terabyte size data files. They demonstrated that the proposed per-posting list indexing strategy was the most efficient indexing strategy, which leveraged a combination of both local machine memory and compression techniques to attack the I/O intensive weakness of map-reduce. Soille et al. (2018) proposed a petabyte-scale platform that consists of hardware and open-source software, distributed file system, and task schedulers for batch processing with containerization of user-specific applications. They proposed an interactive geo-spatial visualization and processing with a series of applications together with performance metrics. Gregory et al. (2019) focused on data search and retrieval from open sources of social networks. They organized a bibliometric study with a focus on information retrieval of user profiles and analysed their behaviour through contextual knowledge, facilitating the design of data discovery systems.

## 2.2　Information Retrieval Framework

Dean (2009) described the requirements for complex design trade-off while building and operating large-scale IR systems to manage by millions of users. The author emphasised on several user-queries with response latency, the size of various corpora-search, the latency and frequency with which documents were updated or added to the corpora, including the quality and cost of the ranking algorithms for retrieval. In addition, the author focused on Google hardware and IR systems and their design challenges. Yang et al. (2015) presented adaptable IR dynamic systems with a sequence of events occurring in artificial intelligence and reinforcement learning. The authors offered IR solutions in a changing environment to learn from past interactions and predict future utilities. In addition, the authors presented advances in IT interfaces with personalization and ad-display demand models through which users intelligently and contextually responded to IR systems in real-time.

Behnert and Lewandowski (2017) described an information retrieval framework for online public access catalogues stored in digital library information systems, demonstrated web search features from heterogeneous library data sources. However, keeping in view the heterogeneity different elements of library information systems posed connectivity challenges and compromised the information quality. Kumar et al. (2016) have developed a framework for data centres to retrieve information, acknowledging the redundancy, security and integrity. The data centres were typically geographically located and characterised as voluminous in the cloud-computing environment. Domain experts further facilitated the information retrieval framework to upgrade the tools in Big Data scale. Yue (2011) described machine-learning techniques that could help resolve the complexity involved in information retrieval models. Further, the author asserted that the learning framework could simplify the overall development process, advocating structured prediction and interactive learning, realizing the feature-rich retrieval models.

Seyler et al. (2018) proposed a framework for leveraging the contextual information for which users, documents and contextualized manuscripts were considered as heterogeneous objects. Graphical embedding was created to learn the objects in the same semantic vector space. Jung (2007) described the heterogeneity of web information spaces, creating uncertainty among search engine users. The author proposed a mediator agent system to estimate the semantics of unknown spaces by learning the fragmented gathers and applying to crawlers. However, the integration of contextual spaces and their semantics were missing due to poorly associated tools. Further, the detachment between contextualisation features and tools was an added challenge while managing complex queries and information search.

Hernandez et al. (2007) explored thematic views of users' specific data motivating semantically indexed contexts and their ontology descriptions. Though the approach has benefit to the semantic representation of contexts, flexibility and adaptability challenges persist. Liu et al. (2010) computed the rank of web pages using real browsing behaviours of web users. The authors described the issues and challenges of the hypergraph link as incomplete and inaccurate when calculate page rank. Instead, they used the real behaviours of the web users. Qin et al. (2010) conducted experiments on benchmark datasets, suggesting practical algorithms for the proposed framework to optimise the IR measures. Zuccon et al. (2013) used a crowdsourcing platform that captured the user interactions, searching their behaviours at low cost with more data within short period times. Still, the approach compromises the quality of information and its search.

Simpson et al. (2014) proposed a practical multimodal solution for indexing and retrieving the images contained in the biomedical articles. Text-based visual representations were used in conjunction with the solutions that significantly improved the retrieval accuracy. Soldaini et al. (2016) investigated the utility of bridging the gaps between layperson and expert vocabularies for query classifications from which a supervised classifier is selected with the most appropriate synonym that best fit with the query. Despite precision in terminology presentation, the classifier lacks collaboration with complex queries. Koopman et al. (2016) present a graphical inference retrieval model that integrates structured knowledge with statistical information and its inference in a unified framework. The analysis suggested when and how to apply the inference for retrieval, including categorization of queries affected by inferences. Though the authors concluded that the inference retrieval method was more effective compared to a general retrieval method, the inference analysis lacked relevant judgement while choosing queries. Tolosa et al. (2017) lately used information retrieval systems at unique stages to speed up computations. The authors proposed a static cache and evaluate its space for query executions and differences to reveal between raw and compressed forms. Karanam et al. (2017) discussed a cognitive model, building a process for information search and enabling the use of semantic spaces. The study concluded with an interpretation of high-domain knowledge in an expert-semantic space irrespective of the knowledge acquisition and lack of refinement process.

Information retrieval and its refinement is much-needed research using an integrated framework. Previous information retrieval researchers have not dealt with a new integrated approach that can cope with complex queries and information needs. We propose a framework to accommodate multiple constructs and models, envisioning the refinement in the search process. We articulate the framework with new artefacts that can search and deliver quality information, minimising the ambiguity of information constrained by various search engines.

## 3   The Existing Issues and Challenges

The information explosion, low information accuracy, improper search outcomes and their management, mismatched human-computer interaction clusters are significant issues, making the information retrieval challenging (Croft et al. 2015). A dynamic and flexible information retrieval system is needed to provide services to users in various KBWE scenarios. In digital ecosystems, other challenges include inadequately integrated domains, systems and their associated data sources, which are briefly summarised in the following sections.

*Integrated search tools*: Specific search tools and functions exist, for example, desktop, music, language-specific searches, and an explicit full-text database with bibliographic searching, in addition to the general-purpose web search engines, such as Google, Bing and Yahoo. However, information seekers must install the tools on their computers, and then match the search tools/functions with their information retrieval needs. The process may involve considerable trial and error and investment in learning a variety of systems. Substantial resources are used such as time, memory, disk space and processing power, in particular, accessing the high-resolution images. On the other hand, integrated search function and data ingest tools can ease the issues of separate search engines and tools by providing a more effective search for a particular domain or field, as compared with the general-purpose search engines and individual search tools.

*The syntax-based search is not necessarily semantic centric*: When the search is syntactical, but not semantic-based, the search results may not be adequately relevant. Web search engines in such cases look for factual similarities between search-terms and the web pages (Arasu et al. 2001). Search engines crawl websites from the Internet and download web pages from various sites. Content of the crawled pages are tokenized and indexed. All tokenized terms are used to create a set of vocabulary, and a term-document matrix can thus be generated. Accordingly, similarities between search-terms and documents are estimated by ranked-based algorithms such as vector space or probability models (Manning et al. 2009). During this process, the semantic characteristics and issues of search-terms may have been incorrectly elaborated. As an example, it is not a surprise when using "UPS" as a search term to retrieve information about the "Uninterruptible Power Supply" that may return irrelevant or ambiguous results such as the "United Parcel Services". Similar is the case with idiom associated with ambiguous "Jaguar".

*Untailored search results*: Wherever contextualized and personalized search is not commonly considered significant, still, most search engines make an effort to return search results based on general-purpose search (Croft et al. 2015). No matter what role a searcher has – a car sales representative, an environmentalist, or a computer technician – if they all search "jaguar", they get the same search results. However, submitting the query, the sales representative might think of the "jaguar" car and not anything else. The environmentalist seeks information about the animal jaguar, whereas the IT technician might think of using Apple's Jaguar as an operating system. The general-purpose search tools thus need improvement to get quality and relevant search results, for example, by creating user profiles. However, Croft et al. (2015)

indicate that using such models does not improve the effectiveness of ranking on its average. On the other hand, local geography-based search, and contextual information extracted from users' interaction with search engines are promising approaches (Croft et al. 2015). Although some search engines provide personalised results, such as Google (Bunz 2009), retrieval and precision are still a concern, as discussed in Croft et al. (2015).

*Enterprise-level search personalisation*: The customisation and personalisation features move conjointly at the enterprise level. Personalisation ensures delivering the content and functionality that match with specific users' needs, what they search for in pursuit of innovative and new terminologies. Netflix is an example, which has established a market with the adaptability of user views and search-terms. Without any clue of queries, most search engines rank results based on the general-purpose search. However, Arnold (2004) cites that the enterprise search is not a web search, although it can manage the indexing of content that resides on the Internet sites. Enterprise search can even support the queries from special classes of authorized users, as in the extranet.

*Integrated holistic enterprise-level search engine*: Enterprise software applications typically control the search functions, making adaptable to integrated search engines. More popular Gmail, Microsoft Outlook and other email services have built-in search functions. Microsoft SharePoint built-in search allows users to search SharePoint pages. Nevertheless, users in an enterprise need to frequently change several portals and explore different types of data relevant to business operations.

As we move forward with new emerging information needs and queries through IR innovations, including for evaluable measures, we articulate a unified and adaptable framework with different artefacts. Semantic knowledge from multiple multimedia sources has significance in analysing the information needs and queries that motivated us to explore alternate IR solutions. With a sheer volume of multimedia information, indexing bears challenges of manual annotations. Our research explores for an integrated framework with improved search algorithms and evaluable measures to facilitate the usability and evaluation criteria. The deficiencies existing in the dynamic KBWE that need a rule-based new algorithmic ISF approach with unified articulations have motivated us to draw research objectives and leverage the KBWE.

## 4   Research Goal, Motivation and Significance

Based on the current information retrieval challenges, we outline the research questions and objectives. Research Questions (RQ) are:

1. How do we design and develop a structured and unified information search framework?
2. What types of algorithms can achieve high precision search results?
3. How do we accomplish the information search through ISF and evaluate the results?

The corresponding Research Objectives (RO) are:

1. Develop ISF that can integrate a variety of search engine tools and applications.
2. Improve search experience using algorithmic approaches.
3. Analyse and evaluate the search results of the ISF.

The goal of the research is to design and develop a framework to search for unified information and validate the given model through a series of experiments with evaluable measures. Another research target is to extract structured textual data that can be shared between different search engines/tools and design the type of framework needed to explore such fused information. Increased search efforts and outcome of terms searched in various contexts is the motivating factor for investigation of the new framework. Activation, persistence and intensity of search terms and their findings usable by type of users motivate us to develop new search articulations. The integrated methodological approach can minimise the ambiguity involved in exploring further information and knowledge acquisition through interfaces and desktop integration.

## 5 Research Methodology

As described in Research Objective 1, we adopt a design science approach, with the methodology motivating a rigor in data science, based on empirical evidence. As a part of design science, we take the support of the approach used in Vaishnavi and Kuechler (2007) and Baskerville et al. (2015), construct coherent and intelligible artefacts within an integrated framework. In pursuance of Research Objective 2, we utilise various algorithmic approaches to build search tools. As a part of empirical research, we propose to analyse the experimental data relevant to search terms and corresponding information needs from multiple search engines and present them for their performance.

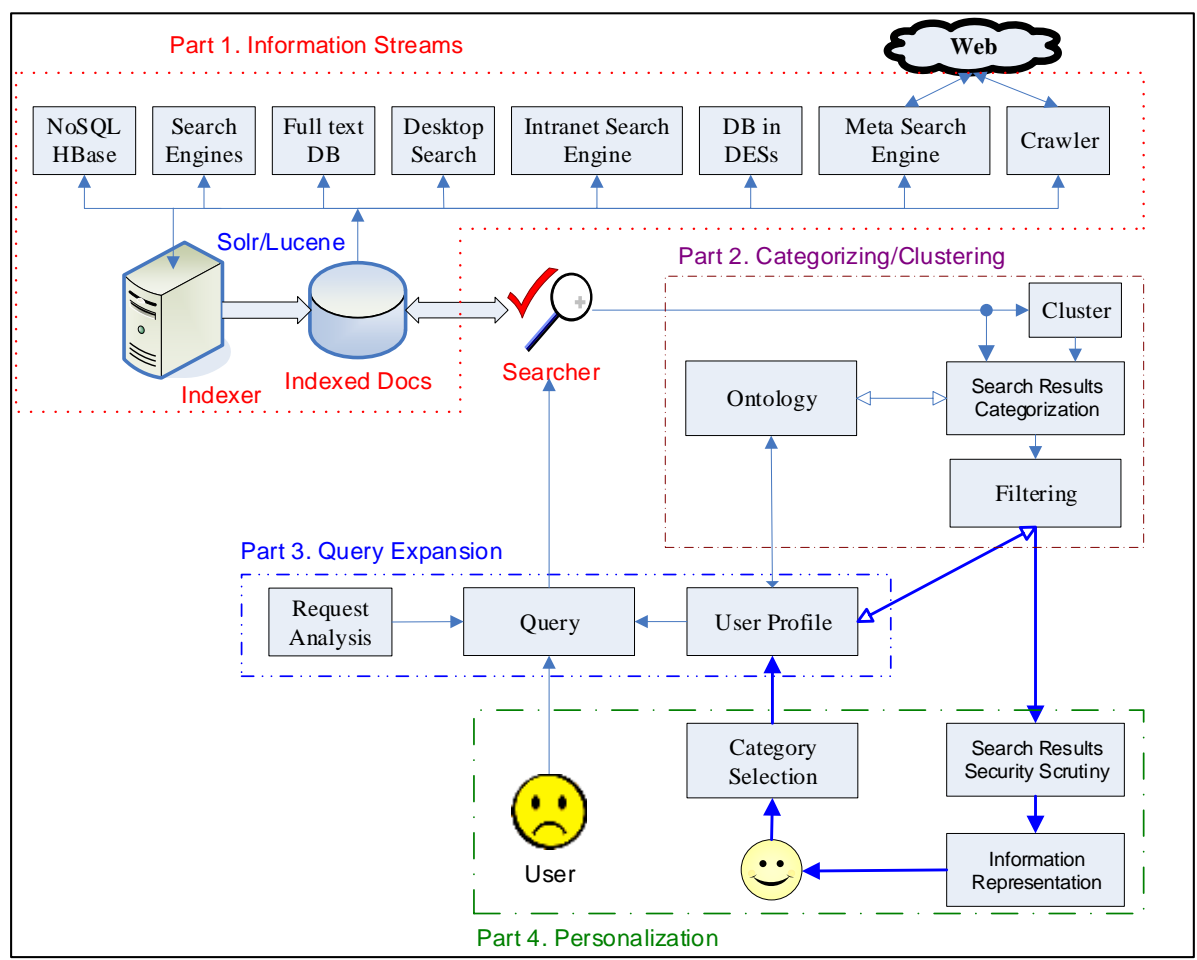

*Figure 1: An Integrated Search Framework (ISF)*

As per Research Objective 3, we carry out various experiments and evaluate the performance of ISF in terms of precision, recall and F1 scores. In this context, we bring the concept of Integrated Search Framework (ISF), as illustrated in Figure 1, put forward with KBWE representation, and articulated with various artefacts. The artefacts are detailed in the following sections.

## 5.1  Information streams

When users submit search requests, search-terms match with information from diverse sources including the Internet, intranets, full-text databases, databases of digital ecosystems and personal desktop computers. Search results are categorized by SVMs or Naïve Bayes based on the ontologies associated with web ecosystems. Search results are also clustered by example-based algorithms such as KNN to further facilitate user to disambiguate semantic features of the search-teams.

The search processes are described in the following sections, with the functionality of each component, as explained in Figure 1. The framework is divided into four main parts, namely, information streams and sources; query or search-term expansion; search results categorizing, clustering and filtering; and personalised search results' representation.

It is worth mentioning that ISF is independent of any specific platform. It can be deployed on an intranet computer cluster with Hadoop Ecosystem, as we experiment on our big data lab's Hadoop cluster. It can also be deployed on a public cloud, which provides IaaS that supports distributed computational and storage capability. It can also be deployed on a hybrid cloud that can provide enhanced security. We can keep sensitive data in the private cloud while the crawled public-accessible data can be stored in a public cloud. Deployment of ISF is transparent to end-users, and from a business perspective, a hybrid cloud model is appropriate for most application scenarios, but not limited to any specific solutions. In short, no matter which platform an organization adopted for its IT infrastructure, ISF can easily be deployed.

*Search process*: The Research Objective 2 motivates us devising new algorithms for retrieving information on the Internet. A programmable script is designed as a crawler for browsing the WWW. Currently, the crawlers emerge with large-size scripts, complexity and rapid growth of the WWW. With the result, the page selection, importance, recency and refresh options create many operational challenges, including limitation of network bandwidth and disk space needed to run the ISF. Other constraints are the rapid update of web content which can make crawled pages obsolete, and limited coverage of web content of a crawler has the potential of missing relevant data from the Web (Baeza-Yates et al. 2005).

As a consequence, in addition to regular page refresh, crawlers should focus on the effectiveness of crawling. Usually, a crawler starts with an initial set of URLs that are placed in a queue and are prioritised. A URL is selected based on specific ordering strategies. The crawlers download web pages, extract URLs from the downloaded pages, and put the new URLs in the queue, expanding into (crawling) relevant websites. This process is repeated until crawlers decide to terminate. Prioritising the URLs in the queue and setting the stop conditions are both related to estimating, or measuring the relevance of the URL content to the semantic need of digital web ecosystems.

PageRank (Brin and Page, 1998) is used to evaluate the relevance of Web pages. Supposed that at a given time, one Web surfer may follow a link with fixed probability $\delta$, then the probability of selecting a page uniformly at random and jumping to the page is $1 - \delta$. If the user can browse the Web quickly and tirelessly, the value of PageRank $\gamma(\alpha)$ is an estimator of the probability that the surfer visits page $\alpha$.

$$r(\alpha) = \delta \cdot \left( \sum_{\beta \to \alpha} \frac{r(\beta)}{o(\beta)} + \sum_{\gamma \epsilon \Gamma} \frac{r(\gamma)}{N} \right) + (1 - \delta) \cdot \sum_{\alpha \epsilon \Phi} \frac{r(\alpha)}{N}$$

Where $\beta \to \alpha$ is the set of all pages that point to page $\alpha$, and o ($\beta$) is the total number of out links from page $\beta$. $\Gamma$ is the set of all sink pages, which also contribute its page rank to $\alpha$. Further, since the choice of $\alpha$ is arbitrary, so we have $\sum_{\alpha \epsilon \Phi} r(\alpha) = N$, thus

$$r(\alpha) = \delta \cdot \left( \sum_{\beta \to \alpha} \frac{r(\beta)}{o(\beta)} + \sum_{\gamma \epsilon \Gamma} \frac{r(\gamma)}{N} \right) + (1 - \delta)$$

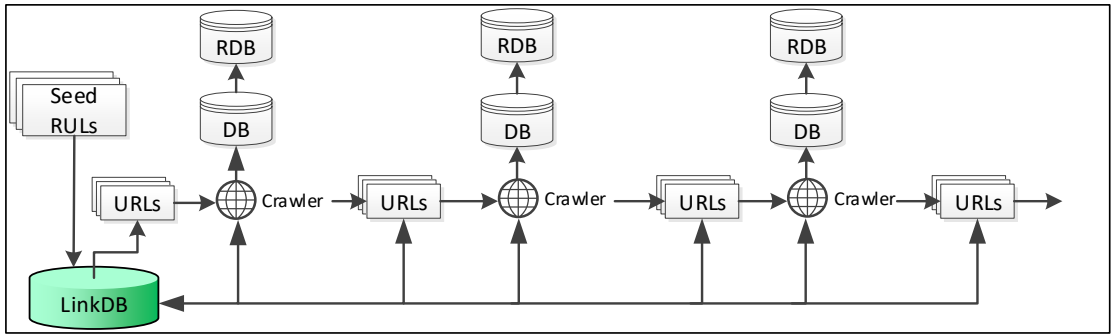

*Figure 2: The crawling strategy for the ISF*

As illustrated in Figure 2, the proposed crawling strategy consists of the following stages. Firstly, a set of seed URLs, which contains all relevant websites picked up by ecosystem domain experts, is created and stored in a URL link-database (LinkDB). The set of URLs are then injected in the URL list (URLs, also URL Frontier). The seed URL list is adjustable based on users' business requirements. For example, an environmental research institute can put only a list of websites related to their research interest into the seed URL list, such as www.environmentalscience.org/sustainability and environmentalprofessionalsnetwork.com/. With the seed URLs injected into the URL frontier, a crawler downloads all Webpages in the URLs and stores the extracted content into a database (DB). A search engine indexes the content after the documents are pre-processed (stop word removing, stemming, named entity extraction). At the same time, URLs contained in the downloaded webpages are extracted and then put into LinkDB after duplicates and irrelevant URLs are removed by applying a filtering strategy, which preserves only pages that are crawled from the domains included in the seed URLs. For all the downloaded webpages in the DB, some of them may not be as relevant as others. Therefore, a document categorisation algorithm such as Support Vector Machines is trained by using existing documents stored in the ecosystem as a training data set. Only the relevant documents moved from DB to the relevant database (RDB) are indexed, where the PageRank calculates the relevance of the documents.

Following is the pseudo-code and a brief explanation for the crawling algorithm with PageRank value priorities as described in Figure 3.

1. Load in seeds URLs and connect to LinkDB
2. extract URLs that can be crawled in next crawling round

3. content parse and process
4. repeat.

After applying relevant filtering algorithms, the hardware storage requirement can be reduced dramatically. In our big data lab (with six servers plus two edge nodes, total 16TB HD, 416GB RAM), we have crawled over 10 million news articles in an Hadoop ecosystem, and the total hard drive used for the crawled pages is only about 140GB in total. Therefore, ISF's filtering algorithms and configurations enable us to keep all relevant data in our private DB.

*Search aggregators*: A metasearch engine is a system that provides unified access to several existing search engines (Meng et al. 2000). The aggregator in ISF is based on the following considerations: 1) single search engine's processing power may not scale to the big increase and virtually unlimited amount of data; 2) it is hard or even impossible for a single search engine to index all the data on the Web and keep it up to date; and 3) some 'deep web' sites may not allow their documents to be crawled by external websites, but allow their documents to be accessible for respective search engines.

```
ISF Crawlor;
    Initilising:
        Starting: load in seed URLS from a text file and put it into linkDB;
        Initialising PriorityQueue with LinkDB;
    Crawling:
    While(PriorityQueue not empty) {
        Read URLs from PriorityQueue;
        Distribute URLs to parallel crawler's waiting queue;
        Foreach parallel crawler: {
            If(waiting queue not empty) {
                Fetch content from URL in the waiting queue;
                Put fetch content into DB;
                Remove fetched URL from waiting list;
            }
        }
    Parse fetched content from DB;
    extract URLs and put them into LinkDB;
    Run PageRank algorithm on LinkDB;
    Remove low PageRank value URLs from LinkDB;
    index content into Solr;
    Stop condition satisfied ?
        Yes: stop;
        No:
            updating PriorityQueue with URLs in LinkDB,
            GOTO Crawling;
```

*Figure 3. Crawler algorithm in ISF*

The conceptual architecture of the meta-search engine is illustrated in Figure 4. Meta-search starts with initializing the user query and selecting a set of suitable databases (coupled with search engines) by the database selector. The document selector chooses the number of documents that are to be crawled from the component search engines. The local similarity threshold can also be used to limit the documents retrieved from the component search engines. Query dispatcher connects the server with each of the selected search engines and passes the query to them. Results manager integrates crawled results from search engines into a single ranked list and renders it to the user (Meng et al. 2000).

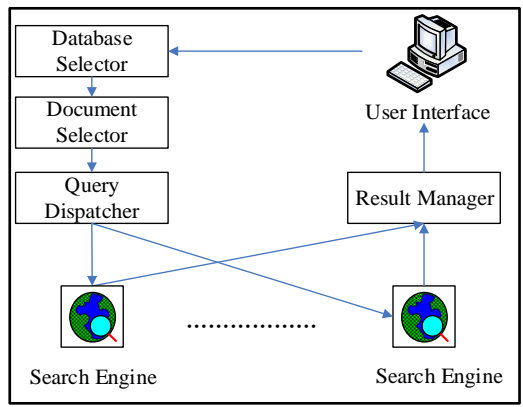

*Figure 4. Meta-search engine structure (Meng et al. 2000)*

*Reprising databases*: A Relational DataBase Management System (RDBMS) typically manages the existing databases in an enterprise that can be re-used by ISF in the KBWE contexts. In addition, the RDBMS manages the metadata and other security-sensitive data, which usually provides security and integrity management (Elmasri and Navathe, 2016). In ISF, RDBMS is one of its data sources and constructive components. User queries are submitted to RDBMS as well as part of the integrated search platform for structured data searching together with other extracted relevant information from all data sources. The retrieved results from the RDBMS are then presented to the categorisation/clustering module for further processing, as illustrated in Figure 1. In addition, the extracted entities such as title, gender, person name, address, email address, organisation, telephone number, and mobile phone number from text documents are stored in the RDMBS. These extracted entities are used to match records of RDBMS and, thus, a connection between structured data in RDBMS, and unstructured/semi-structured text data can be established, and verse vice. Users of ISF can search for data by providing a Google-like search term and can inspect data in the RDBMS as well. NoSQL database such as HBase data can also be ingested into ISF. HBase (http://hbase.apache.org) is a schema-less NoSQL-style data storage designed for large scale, over billions of rows and millions of columns, random read and write operations of sparse data. The data in DB and RDB, as shown in Figure 2, is stored in HBase in ISF. Solr's Java API package Solrj is used to index data stored in HBase. Pseudo-code is presented in Figure 5, indexing HBase data into Lucene/Solr Cloud, an open-source search service that indexes data in ISF and response queries for the indexed data with blazing-fast response speed. ISF utilizes SolrCloud to provide high availability, scalability and fault tolerance. Hadoop and SolrCloud enable ISF the ability to linearly scale out its computations and storage capability by merely adding more working nodes to the cluster when more data storage and computations are required.

```
Import solrj packages;
    Import Hadoop hbase java packages;
    Connect to solr as solr client;
    Connec to hbase ;
    Scan a hbase table and get results;
    For each row in results
        Extract data in htable cell and write the data to solrj documents;
    Solrj index documents;
```

*Figure 5. Pseudo-code used for indexing*

*Desktop search*: As indicated by Barrows and Traverso (2006), Google, Microsoft, Yahoo! and other major search engine players are providing free downloaded desktop search solutions. As the storage allows hundreds of gigabytes of data and beyond, the desktop search can significantly make better user productivity. In these models, indexing or tagging can enable users to access information through dynamic integrated navigational retrieval systems. They can act as agents to return pointers or links to the desired information.

*Intranet engine*: Searching for information on the intranet is rather a daunting task, which is addressed by search tools development companies, such as Google and Thunderstone. Because the performance of intranet search engines differs dramatically, ISYS Search Software (www.isysdemo.com) suggests that features should be considered carefully. However, in ISF contexts, the intranet engine selection is the user' choice.

## 5.2 Search results categorisation and clustering

Knowledge-based hierarchical ontologies are considered for categorising and interpreting a considerable amount of information and its semantic content. Another method is to arrange the itemized information into different clusters according to their similarities. In ISF, these two approaches are combined to leverage the advantages of both tactics, as suggested in Chau and Chen (2008).

*Fine-grained ontologies*: The ontologies are created articulating various artefacts in the ISF to improve the search process, categorising the search results based on semantics. The search results are further filtered based on the user selection and accordingly classified under the selected category for description and presentation. In ISF, the Open Directory Project (ODP, www.odp.org) is employed as an ontology to present the Web knowledge structure. The semantic characteristics of each category in the ODP are manifested by a category-document that includes the topic of the ODP category, the description of the type and a list of submitted Web-pages (composed of the title of the Web-pages and a brief description of each of the submitted Web-pages) under this category (Zhu et al. 2018).

*Categorising search results*: Text categorisation is the problem of assigning pre-defined categories to free-text documents (Croft et al. 2015). In ISF, search results are categorised based on the ODP as a lightweight ontology. The category-documents in the ODP are employed as a training data set and an example-based categorization algorithm, which utilizes a vector space model (Manning 2009), is described as following.

Let $d_j = \{w_{1,j}, w_{2,j}, \ldots w_{T,j}\}$ be the *j*-th category-document, where *T* is the total number of vocabularies in ISF, and $q = \{w_{1,q}, w_{2,q}, \ldots w_{T,q}\}$ is the search item; $w_{i,j}$ is the tf-idf (term frequency-inverse document frequency) weight of *i*-th term in *j*-th document, $w_{i,q}$ is the tf-idf weight of *i*-th term in the search item. The similarity between *q* and *d_j* is estimated by the cosine value of the angle Θ of the two vectors:

$$sim(d_j, q) = \cos(\theta) = \frac{d_j \bullet q}{|d_j| \times |q|} = \frac{\sum_{i=1}^{T}(w_{i,j} \times w_{i,q})}{\sqrt{\sum_{i=1}^{T} w_{i,j}^2 \sum_{i=1}^{T} w_{i,q}^2}}$$

The similarities between *q* and the *d_j*, *j* = 1, 2 … N (where N is a total number of *category-documents* in ISF) are ranked/sorted in their descending order. For top K ranked category-documents, suppose their corresponding ODP category is C = {$c_1$, $c_2$… $c_K$}, *q* is assigned to the category selected from C by dominant majority voting algorithm, as shown in Figure 6.

```
if (m > k/2)
        Assign the document the category decided by the majority members
else if (m <= k/2) {
        if (top ranked member is among the majority group)
                Assign the document the category decided by the top
                ranked member
        else
                Assign the document two categories, one is the same as
                the top ranked member, another is decided by the
                majority member.
}
```

*Figure 6. Majority voting algorithm for text classification*

*Clustering*: Text clustering aims at assembling documents that are related among themselves and satisfy a set of features. It can be used to expand a user query with new and related index terms (Croft et al. 2015) and to facilitate users to browse the retrieved results. Many clustering algorithms are available, such as the *K*-mean clustering algorithm and Fuzzy *C*-Means. In ISF, *K*-mean is chosen to cluster returned search results. Two essential issues of K-means are how to decide a proper K and how to select the original K cluster centres. Since the search results are categorized based on the ODP category, the number of the first level categories under which search results are assigned is a suitable candidate for *K* in ISF. Meanwhile, the returned search result $r_k$ that is most similar to a candidate category $c_j$, is assigned as the centre of a new cluster, and as a consequence, we can select *K* <= *C* clusters for search results clustering where *K* is the number of generated clusters and *C* is the total number of top-level ODP categories. Cosine similarity (as described above) is utilized to estimate the "likelihood" among neighbouring groups.

*Filtering*: Google and Facebook introduced personalisation features with improved algorithms that filter information as per user requirements and motivate the filtering process. We further analyse filtering processes to show how the personalisation can be linked to filtering techniques without any bias of human and computer interaction. Search result filtering is decided in ISF by two factors: one is the user's selection of an existing category of the ontology; another factor is the pre-built user profile that is to be discussed in the next section. With a user choosing several categories, only search results categorised under the categories are presented to the user, and other information is filtered out. However, even if the user does not select any pre-defined ODP categories, the search results are filtered as well by default based on a pre-built user profile. Search results are compared with the features in user profiles and re-ranked accordingly to ensure only results having similar features described by user profiles are presented to the users.

### 5.3   Expanding and analysing queries

*User profile and personalisation*: A user profile is a reference ontology in which each concept has a weight indicating the user's interest in that concept (Croft et al. 2015). An information space of the ODP (Pitkow et al. 2002) is used to represent user models. As suggested in Dolog and Nejdl (2003), the user model combines two proposed standard learner profiles, IEEE Personal and Private Information and IMS Learner Information Package (LIP) to express the features of a user. The precision and recall of metasearch engines are improved through mining association rules that reflect the user' past search behaviour. ISF cuts off the ODP knowledge hierarchy from the second level to obtain 573 topics and uses these topics to represent users' search interests that are represented by <*topic, weight*> tuples. The user- profile is initialised by asking users to assign a weight (integer) to an existing topic to indicate how import the topic is. Users are allowed to choose any number of interesting topics. To map a user search interest into these topics, for each search result $r_i$ visited by a user, let $c_i$ is the topic, as detailed in Section 2, the corresponding weight of $c_i$ in the tuple is increased.

*Query and request analysis*: Query augmentation and result processing are two primary uses of user profiles. In the ISF scenarios, after a user selects an ODP topic, the query takes as an alternative augmentation to user profiles, which allows the user to re-submit the query $q+ = q \cup \{t_k, k = 1, 2 \ldots K\}$, where $q$ is the query submitted by the user, $t_k$ is the term selected using

$$\lambda^2 = \frac{N[P(t_k,c_i)P(\bar{t}_k,\bar{c}_i) - P(t_k,\bar{c}_i)P(\bar{t}_k,c_i)]^2}{P(t_k)P(\bar{t}_k)P(c_k)P(\bar{c}_k)}$$

Where $N$ are the 573 topics in the user profile, and these topics are now represented by *category-document*. $P(t, c)$ is the joint probability of term $t$ and category $c$; $\bar{c}$, $\bar{t}$ indicate that $c$ or $t$, respectively, do not appear in that category. $K$ is the number of terms determined by the confidence $P(\lambda^2 > 10.83) < 0.001 = 99.9\%$ that the assumption of independence of the term $t_k$ and $q$ can be rejected (Manning et al. 2009).

## 5.4 Representing personalised search results

*Information representation*: In the ISF, search results are extracted from full-text databases, intranet, Internet, traditional databases and desktop searches - all results are integrated into one coherent information representation. Users of ISF can choose which data sources are to be included during the search process, thus providing the flexibility to access data sources to satisfy their information needs.

*Search results from security scrutiny*: The component in the ISF performs a search result scrutinizing security feature. A security scrutiny task concerns "who is allowed to update a piece of metadata and view a particular piece of metadata about a document or know that the document exists at all" (Barrows and Traverso, 2006). We apply Apache sentry with Solr that makes ISF with the ability to fine-grained control of access of data indexed by SolrCloud.

## 6 Experimental Results

Järvelin (2007) has discussed two versions of IR approaches, one with the lab version on the design of retrieval models with refinements in evaluable query results, and the other relevant to cognitive IR approach with object-oriented data entities, structures and their relationships. In our research, we conduct experiments from WWW through KBWE using a number of queries and information needs. In this section, we use processes, methods, artefacts, and algorithms to retrieve information that is accountable to new knowledge from ISF and comparable with other search engines. For ascertaining the research purpose and evaluating the search process through artefacts and articulations of ISF, the data are generated through experiments with evaluable measurements. A pragmatic approach is adopted with initial experiments done to establish an information retrieval method and a search process mechanism as described in Zhu et al (2018). Additional experiments are carried out for testing the ISF and its search precision through different queries, information needs, and validation of relevant judgements as interpreted in terms of the cut-off levels P@5 and P@10.

*Evaluable measurements*: The two widely accepted measurements of information retrieval effectiveness are precision and recall. Recall measures the ability of an information retrieval system to retrieve all relevant documents, whereas precision measures the ability of an information retrieval system to extract only relevant items. For text categorisation purposes, the two measures are defined as (Manning et al. 2009):

$$recall = \frac{categories\ found\ and\ correct}{total\ categories\ correct}$$

$$precision = \frac{categories\ found\ and\ correct}{total\ categories\ found}$$

For the evaluation process, search engines are made available for retrieved documents in a ranked list according to the degree of relevance of the documents to a given query (Alonso and Mizzaro, 2009). Users examine the ranked list starting from the top document. The recall and precision measures vary since the users proceed with an examination of the returned set.

In addition to the overall precision recall measure, a precision versus recall curve based on eleven standard recall levels is usually employed to evaluate the ranked list (Croft et al. 2015). Another measurement is the precision at a given cut-off level. A cut-off level is a rank that defines the retrieved set. For example, a cut-off level of 10 represents the top ten retrieved documents in the ranked list. If seven out of ten of the returned documents is relevant, the precision at cut-off level ten (P@10) is 7/10 = 0.7 = 70%.

*Experiment description*: A special search browser (SSB), which is a component of ISF, has been developed to categorise Web search results from Yahoo! under various categories of the ODP (Zhu et al 2018). Five search-terms with general or ambiguous meanings are selected, as shown in Table 1. For each search-term, 50 search results are retrieved by utilizing the Yahoo! Search Web Service API. The returned search results are presented to human judges to perform relevance judgment. The relevant judgment results are summarized using a weighted schema. A final binary decision is made regarding whether a returned search item is relevant to the specified information need or not. Based on the relevance judgment results, the standard 11 recall-precision curves are drawn for each search-term of the search results of Yahoo!, and of the categorised results of SSB, as shown in Figure 7a. P@5 and P@10 of the search result sets are presented in Table 2.

| Query | Information needs |
|---|---|
| Clinton | The American president William J Clinton |
| Ford | Henry Ford, the founder of Ford motor company |
| Health | State of physical, mental, and social well-being |
| Jaguar | Information about the animal "jaguar" |
| UPS | Information about "uninterrupted power supply" |

*Table 1. Search Terms and Information Needs*

|  | P@5 | P@10 | Average |
|---|---|---|---|
| Yahoo | 46.7 | 42 | 44.4 |
| SSB (ISF) | 85 | 70 | 77.5 |
| Improvement | 38.3 | 28 | 33.2 |

*Table 2. P@5 and P@10 of Yahoo and SSB*

| Query | Google | | Bing | | Lycos | | MS Live | |
|---|---|---|---|---|---|---|---|---|
| | P5 | P10 | P5 | P10 | P5 | P10 | P5 | P10 |
| Clinton | 40 | 40 | 20 | 10 | 40 | 40 | 0 | 0 |
| Ford | 20 | 30 | 40 | 20 | 20 | 20 | 20 | 20 |
| Health | 100 | 100 | 100 | 70 | 100 | 80 | 60 | 40 |
| Jaguar | 20 | 40 | 20 | 10 | 20 | 40 | 20 | 10 |
| UPS | 20 | 10 | 0 | 0 | 20 | 10 | 20 | 20 |
| Average | 40 | 44 | 36 | 22 | 40 | 38 | 20 | |
| | 42 | | 29 | | 39 | | 21 | |

*Table 3. P@5 and P@10 of Google, MS Live Search, Bing, and Lycos*

The data documented in Tables 2 and 3 summarize the relevant judgement results and the summary of estimation. Following the macro-averaging style (Manning et al. 2009), while drawing the standard 11-point recall-precision curve, the precision $p_j$ at recall level $j$ is calculated by:

$$p_j = \frac{1}{N} \sum_{i=1}^{N} p_{i,j} \quad j = 0, 1, ..., 10$$

$N = 5$ is the number of queries in the experiment, $p_{i,j}$ is the precision of the *i*-th query at *j*-th recall level. The overall precision is calculated by:

$$p = \frac{1}{11 \times N} \sum_{j=0}^{10} \sum_{i=1}^{N} p_{i,j}$$

*Relevance judgment*: Relevance judgment is inherently subjective (Mizzaro, 1997). For easing the issue of subjectivity, we organized five judges from our university to conduct relevance judgement in the research, as discussed in Zhu et al (2018). The human judges know nothing about the categorised results of the ISF. According to the relevant judgement results of five judges, a final binary decision was reached for each returned search item of Yahoo API. The experimental results in Figure 7a and Table 2 of ISF are based on 50 search results for each of the five search terms of Yahoo!. That is, the ISF has categorised Yahoo's 50 results into different ODP categories. In this context, the comparison between Figure 7a and Table 2 is a direct comparison. Because ISF does not categorise the search results of the rest of four search engines, as in Figure 7b, for the reason they do not have corresponding search APIs as provided by Yahoo!, and in such sense, the comparison may not be considered straightforward. However, the indirect comparison also reveals that without applying the proposed search strategy, the performance of the search engines is relatively unsatisfactory regarding the recall-precision curve measures and the P@5 and P@10 evaluations.

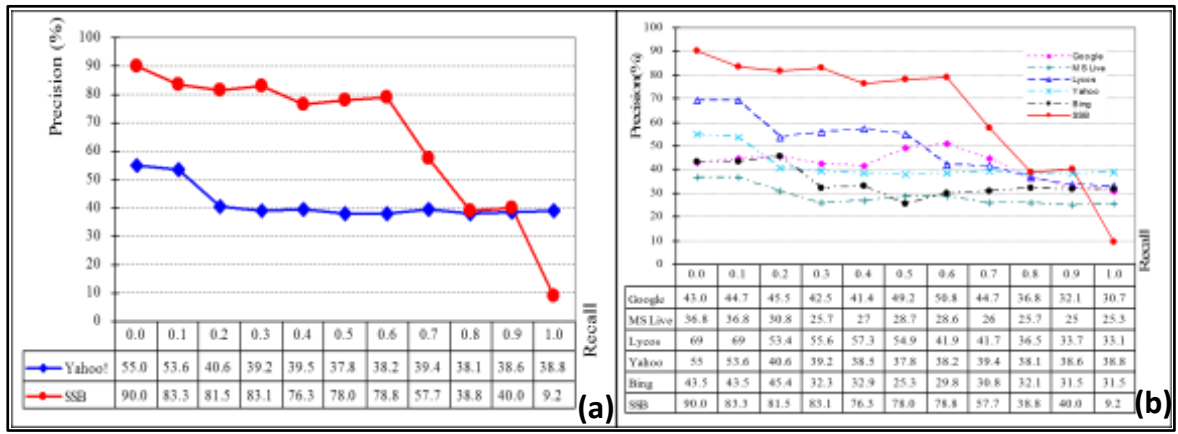

*Figure 7 (a): Average recall-precision curves of Yahoo! and ISF categorized search results over the five search-terms (Clinton, Ford, Health, Jaguar, UPS) (b) Average recall-precision curves of Yahoo!, Google, MS Live Search, Bing, Lycos, and the categorised results of ISF[1]*

To further test the ISF, we have added five more queries with information needs as shown in Table 4. The search results for P@5 and P@10 are presented in Table 5.

| Query | Information Needs |
|---|---|
| susan dumais | the researcher Susan Dumais |
| world war 2 | history related to World War 2 |
| graphic design | the art and practice of graphical design |
| Jokes | the funniest jokes |
| Time zones | time zones of the world |

*Table 4. Five more Search Terms and Information Needs*

| Query | Google | | Yahoo | | Bing | | AOL.Com | | Baidu | |
|---|---|---|---|---|---|---|---|---|---|---|
| | P@5 | P@10 | P@5 | P@10 | P@5 | P@10 | P@5 | P@10 | P@5 | P@10 |
| Susan Sumais | 100 | 90 | 80 | 70 | 80 | 70 | 80 | 70 | 60 | 60 |
| World War 2 | 100 | 100 | 100 | 90 | 100 | 90 | 100 | 100 | 100 | 80 |
| graphic design | 60 | 50 | 40 | 60 | 60 | 80 | 20 | 60 | 100 | 100 |
| jokes | 80 | 70 | 80 | 80 | 80 | 80 | 80 | 80 | 100 | 90 |
| time zones | 80 | 80 | 100 | 100 | 100 | 100 | 80 | 80 | 100 | 100 |
| Average | 84 | 78 | 80 | 80 | 84 | 84 | 72 | 78 | 92 | 86 |
| | 81 | | 80 | | 84 | | 75 | | 89 | |

*Table 5. P@5 and P@10 instances of Google, MS Live Search, Bing, Yahoo, AOL.com and Baidu*

To further verify the robustness and adaptability of ISF, we have added another five more queries as represented in Tables 6 and 7.

---

[1] SSB is an earlier version of ISF so we kept the name of SSB unchanged in this figure.

| Query | Information Needs |
|---|---|
| sports | sport news |
| soccer league | names of soccer league teams |
| viruses | description of viruses |
| Coronavirus | Coronavirus status |
| environment | the environment species surviving |

*Table 6. Five more Search Terms and Information Needs*

| Query | Google | | Yahoo | | Bing | | AOL.Com | | Baidu | |
|---|---|---|---|---|---|---|---|---|---|---|
| | P@5 | P@10 | P@5 | P@10 | P@5 | P@10 | P@5 | P@10 | P@5 | P@10 |
| sports | 80 | 60 | 100 | 90 | 80 | 70 | 100 | 100 | 80 | 90 |
| soccer league | 80 | 100 | 80 | 70 | 80 | 60 | 60 | 70 | 80 | 90 |
| viruses | 100 | 100 | 80 | 70 | 80 | 70 | 20 | 60 | 60 | 70 |
| Coronavirus | 80 | 90 | 100 | 80 | 60 | 50 | 40 | 60 | 60 | 70 |
| environment | 80 | 70 | 60 | 50 | 50 | 40 | 40 | 60 | 60 | 70 |
| Average | 84 | 84 | 84 | 72 | 70 | 58 | 52 | 70 | 68 | 78 |
| | 84 | | 78 | | 64 | | 61 | | 73 | |

*Table 7. P@5 and P@10 instances of Google, MS Live Search, Bing, Yahoo, AOL.com and Baidu*

A summary overview of the experimental data:

- As shown in Figure 7a, the overall average *precision* of the 50 search results of Yahoo! is 458.8 / 11 = 41.7%.

- The overall average *precision* of the categorized results is 716.7 / 11 = 65.2%

- The average improvement on the *precision* of the categorised results is 65.2% – 41.7% = 23.5 (%)

- Table 2 demonstrates that ISF improves P@5 and P@10 by 38.3% and 28.0%, respectively, with an average improvement of 33.2%.

- To lessen the issue of subjectivity (Mizzaro 1997), the relevant judgement is further analysed by three judges. Figure 8 and Table 5 demonstrate the recently conducted search results from multiple search engines.

- Table 5 illustrates Bing and Baidu search engines perform better retrieval than Yahoo! and Google, all outperform AOL, for added queries described in Table 4.

- Table 7 shows Google exhibits better results compared with other search engines.

# 7   Discussions

*Search results of ISF, Google, Bing, MS Live and Lycos for queries in Table 1:*

Figure 7b and Table 3 reveal that 1) the ISF outperforms other search engines regarding averaged precision based on standard 11 recall-precision curves and P@5 and P@10 in the experiment. 2) The performance of Google and Lycos is nearly the same; MS Live Search and MS Bing perform relatively weak, as both the recall-precision curve and P@5 and P@10 demonstrated. 3), Lycos performs better by the measure of the recall-precision curve, but only better than MS Live Search and Bing when evaluated by P@5 and P@10 criteria.

*Search results of ISF, Google, Bing, Yahoo, AOL and Baidu based on all 15 queries:*

To verify the effectiveness of ISF, Recall-Precision curve is drawn based on the search results of all fifteen-search terms given in Tables 1, 4 and 6 and, as shown in Figure 8.

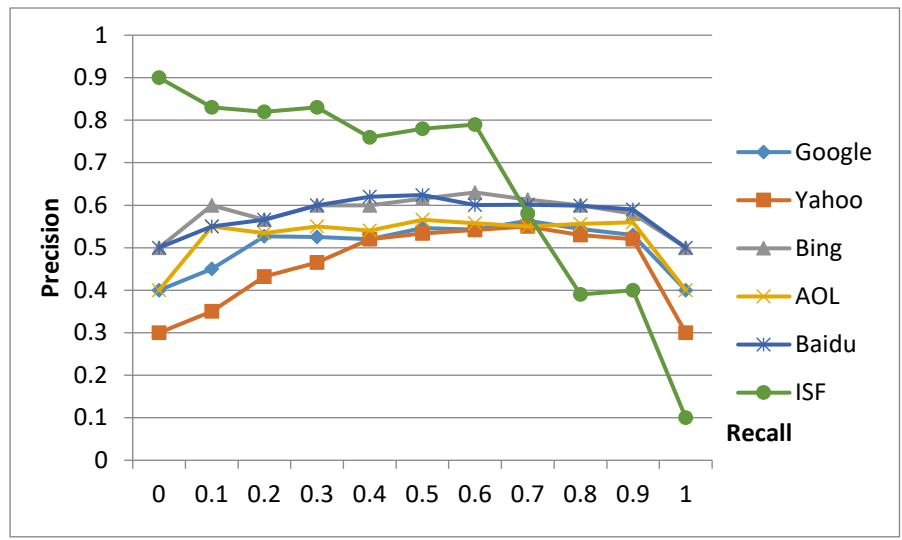

*Figure 8: Comparative performances between various search engines*

Figure 8 compares search engines by recall-precision curves. Queries and their precision instances are given in Table 3 and Table 5 are combined in the schematic view representation in Figure 8 to assess the search engine performances. The ISF emerges as a performing tool, with a better search than any other investigating engines considered in the study. The recall-precision curves for all search engines show steady trends and patterns, compared with ISF. The ISF exhibits high precision at lower recall values, but its precision decreases at higher recall values. The curves of other search engines exhibit a concave-down shape, implying that the search precision and information qualities respond to early and late recall instances. The precision is steady for recall range of 0.3 – 0.9. ISF appears more sensitive to higher recall values. However, we deal with a trade-off between the information needs and queries, because of the broader scope of existing search engines and ambiguities arisen in among complex queries, information needs and interrogative relevant judgements.

The Google, Bing, Yahoo, AOL.com and Baidu search engines could return significant results for all P@5 and P@10 queries for the five new queries in Table 4, but relatively weak for queries in Table 1, except "health", which we give a more general information need. The information need "Clinton" has delivered weak results for all search engines. Syllables "UPS and Jaguar" appear to be ambiguous for all search engines that returned comparatively insignificant results for both queries. As shown in Table 3, although MS Live could return relevant information for "health", for the rest four queries, the formed clusters are in general irrelevant to the specified information needs. For example, when search results of "jaguar" are clustered, all formed clusters are about cars: Jaguar Cars, Jaguar XF, Jaguar UK, Jaguar Dealers, Jaguar Accessories, BMW, Mercedes, and Jaguar X Type. MS Live performed even worse, and we could hardly find relevant information from it.

The Google, Yahoo, MS Bing, Aol.com, and Baidu search engines respectively provide 49%, 50%, 62%, 54% and 59% precision levels based on 11 recall-precision curves when search results from all fifteen queries are summarized. Whereas ISF offers 65.2%, a better precision compared with other search engines. Largely, the framework validates the case studies with variables used in the search results.

## 8  Future Scope and Limitations

The experimental results so far demonstrate that text categorization in ISF improves the precision of Web search results. Implementing the rest of the search framework and conducting a wide range of experiments are ongoing research. Though improved precision is observed, the recall of categorized search results is lower than the search results of Yahoo and other search engines. The reasons for the issue are, firstly one search result is assigned to only one category, even if the second or third ranked category has very close similarity instance with the top one. Another reason is that categorization algorithm utilized in the ISF is not optimal. Combining text clustering and categorization is likely the next research direction, which can improve the recall of the categorized Web-search results.

Since search engine results are dynamic, the compared results from Google, Bing and others are based on search requests when we conducted the experiments and can vary with time. In addition, these search engines can also allow personalized search; therefore, a further comparison of personalized results of search engines with ISF can fetch enormous scope and opportunity in the IR research.

## 9  Conclusions

The information search with ISF is aimed at providing effective information retrieval services leveraged by the KBWE. It is an innovative approach compared with the existing search methods. ISF integrates not only traditional database search (MIS) and Web search engines, but also effectively bringing intranet, desktop and full-text database search tools into the framework. The personalization, ontological search results categorization, clustering and security scrutiny are added features. With the collaboration of empirical research, the experimental data demonstrate that text categorization based on the ISF can improve the precision of Web search results. The algorithms developed in the ISF template have provided high information precision value in comparison with the latest search engines. In terms of accuracy, experience and innovation, the ISF appears promising. The comparison made among search engines offers new improvements with satisfactory search results from the ISF.

## References

Albro, E. N. (2006). Google Mini Is a Mighty Search Tool," *PC World*, https://www.pcworld.com/article/126139/article.html, June 21.

Alonso, O. & Mizzaro, S. (2009). Relevance criteria for e-commerce: a crowdsourcing-based experimental analysis, *Proceedings of the 32nd international ACM SIGIR conference on research & development in information retrieval*, 760-761, Boston, MA, USA — July 19 - 23, 2009, https://doi.org/10.1145/1571941.1572115.

Arnold, Stephen E. (2004). *How Google Has Changed Enterprise Search*. In: Searcher 12, S. 8-17.

Arasu, A. Cho, J. Garcia-Molina, H. Paepcke, A. & Raghavan, S. (2001). Searching the Web," *ACM Transactions on Internet Technology*, 1 (1), 2001, 2-43.

Baeza-Yates, R., Castillo, C., Marin, M. & Rodriguez, A. (2005). Crawling a Country Better Strategies than Breadth-First for Web Page Ordering. *The 14th international conference on World Wide Web*, May 10–14, 2005, Chiba, Japan.

Barrows, R. & Traverso, J. (2006). Search Considered Integral," *ACM Queue,* May 2006, 30-36.

Baskerville, R. L., Kaul, M., & Storey, V. C. (2015). Genres of Inquiry in Design-Science Research: Justification and Evaluation of Knowledge Production. *MIS Quarterly*, 39 (3), 541-564.

Behnert, C. & Lewandowski, D. (2017). "A framework for designing retrieval effectiveness studies of library information systems using human relevance assessments", *Journal of Documentation*, 73 (3), https://doi: 10.1108/JD-08-2016-0099

Brin, S. & Page, L. (1998). "The anatomy of a large-scale hypertextual Web search engine" *Computer Networks and ISDN Systems*. 30 (1–7): 107–117. CiteSeerX 10.1.1.115.5930. https://doi.org/10.1016/S0169-7552(98)00110-X

Bunz, M (2009). "Google extends personalised search to all users". *The Guardian*. Tue 8 Dec, 2009. https://www.theguardian.com/media/pda/2009/dec/07/google-personalised-search.

Chau, M. & Chen, H. (2008). "A Machine Learning Approach to Web Page Filtering Using Content and Structure Analysis," *Decision Support Systems*, 44 (2), 482-494.

Croft, W.B., Metzler, D. & Strohman, T. (2015). *Search Engines – Information Retrieval in Practice*, Pearson Education, Boston, USA.

Dean, J. 2009. Challenges in Building Large-Scale Information Retrieval Systems, Google, *ACM Conference Series, ACM International Conference on Web Search and Data mining, WSDM 2009*, https://pdfs.semanticscholar.org/fc32/72302461b74217662085a8a05a5e500dbf05.pdf

Dolog, P. & Nejdl, W. (2003). Challenges and Benefits of the Semantic Web for User Modelling, In De Bra, P., Davis, H., Kay, J. and Schraefel, m. (eds.) *Proc. of AH2003: Workshop on adaptive hypermedia and adaptive Web-based systems*, Budapest, Hungary, Eindhoven University of Technology, pp. 99-111. Available online at: <http://wwwis.win.tue.nl/ah2003/proceedings/um-1/>.

Elmasri, R., & Navathe, S. (2016). *Fundamentals of database systems*, Hoboken, NJ : Pearson, USA, 2016.

Gartner, (2017). "Insi*ghts From the 2017 Gartner CIO Agenda Report: Seize the Digital Ecosystem Opportunity*," 2017.

Gregory, K. M., Cousijn, H., Groth, P. Scharnhorst, A. & Wyatt. S. (2019). Understanding Data Search as a Socio-technical Practice, *Journal of Information Science*. https://doi.org/10.1177/0165551519837182

Haneef, I., Munir, E. U., Qaiser, G., Hafiz Gulfam, H. & Ahmad, U. (2018). Big Data Retrieval: Taxonomy, Techniques and Feature Analysis, *IJCSNS International Journal of Computer Science and Network Security*, 18 (11).

Hernandez, N. Mothe, J., Chrisment, C., & Egret, D. (2007). Modeling context through domain ontologies, *Information Retrieval Journal* (2007) 10:143–172, https://doi 10.1007/s10791-006-9018-0

Järvelin, K. (2007). An analysis of two approaches in information retrieval: From frameworks to study designs, *Journal of the American Society for Information Science and Technology,* 58 (7), https://doi.org/10.1002/asi.20589

Jung, J. J. (2007). Ontological framework based on contextual mediation for collaborative information retrieval, *Information Retrieval Journal* (2007) 10:85–109. https://doi 10.1007/s10791-006-9013

Karanam, S., Jorge-Botana, G., Olmos, R. & Oostendorp, H. V. (2017). The role of domain knowledge in cognitive modelling of information search, *Information Retrieval Journal* (2017), 20:456–479. https://doi 10.1007/s10791-017-9308-8

Koopman, B., Zuccon, G., Bruza, P. Sitbon, L. & Lawley, M. (2016). Information retrieval as semantic inference: a Graph Inference model applied to medical search, *Information Retrieval Journal* (2016) 19:6–37. https://doi 10.1007/s10791-015-9268-9

Kumar, S. S., Mahapatra, D. P. & Balabantaray, R. C. (2016). Challenges for Information Retrieval in Big data: Product Review Context, *International Journal of Computer Applications* (0975 – 8887), 136 (3), February 2016.

Liu, Y., Liu, T. Y., Gao, B., Ma, Z. & Li, H. (2010). A framework to compute page importance based on user behaviours, *Information Retrieval Journal* (2010) 13:22–45. https://doi 10.1007/s10791-009-9098-8

Manning, C. D. Raghavan, P. & Schütze, H. (2009). *Introduction to Information Retrieval*, Cambridge: Cambridge University Press, New York, NY, USA, 2009.

McCandless, M., Hatcher, E. & Gospodnetić, O. (2010). *Lucene in Action*, 2nd, Greenwich: Manning Publications.

McCreadie, R., Macdonald, C. & Ounis, L. (2012). MapReduce indexing strategies: Studying scalability and efficiency, *Information Processing & Management,* 48 (5), September 2012, 873-888. https://doi.org/10.1016/j.ipm.2010.12.003

Meng, W., Yu, C. & Liu, K. L. (2000). Building Efficient and Effective Metasearch Engines," *ACM Computing Surveys,* 34 (1), 48-89.

Mizzaro, S. (1997). "Relevance: The Whole History," *Journal of the American Society for Information Science* 48, 810-832.

Moore, R., Seedat, Y., & Chen, J. Y. J. (2018). *South Africa: Winning with Digital Platforms*, Accenture, 2018.

Pitkow, J. Schütze, H. Cass, T. Cooley, R., Turnbull, D. Edmonds, A. Adar, E. & Breuel, T. (2002). Personalized Search: A contextual computing approach may prove a breakthrough in personalized search efficiency," *Communications of the ACM*, 45 (9), 50-55.

Qin, T., Liu, T. Y. & Li, H. (2010). A general approximation framework for direct optimization of information retrieval measures, *Information Retrieval Journal* (2010) 13:375–397. https://doi 10.1007/s10791-009-9124-x

Seyler, D., Chandar, P. & Davis, M. (2018). An Information Retrieval Framework for Contextual Suggestion Based on Heterogeneous Information Network Embeddings, *SIGIR '18, July 8–12, 2018, Ann Arbor, MI, USA c 2018 Association for Computing Machinery*. Retrieved from https://doi.org/10.1145/3209978.3210103

Simpson, M. S., Demner-Fushman, D., Antani, S. K. & Thoma, G. R. (2014). Multimodal biomedical image indexing and retrieval using descriptive text and global feature mapping, *Information Retrieval Journal* (2014) 17:229–264. https://doi 10.1007/s10791-013-9235-2

Soille, P., Burger, A., Marchi, D. D., Kempeneers, P., D.Rodriguez, D., Syrris, V. & Vasilev, V. (2018). A versatile data-intensive computing platform for information retrieval from big geospatial data, *Future Generation Computer Systems*, Elsevier, Volume 81, April 2018, Pages 30-40, https://doi.org/10.1016/j.future.2017.11.007

Soldaini, L., Yates, A., Yom-Tov, E., Frieder, O. & Goharian, N. (2016). Enhancing web search in the medical domain via query clarification, *Information Retrieval Journal* 19 (1-2), 149-173 (2016). https://doi 10.1007/s10791-015-9258-y

Tolosa, G., Feuerstein, E., Becchetti, L. & Marchetti-Spaccamela, A. (2017). Performance improvements for search systems using an integrated cache of lists + intersections, *Information Retrieval Journal* (2017) 20 (3):172–198. https://doi.org/10.1007/s10791-017-9299-5

Vaishnavi, V. K. & Kuechler, W. (2007). *Design Science Research Methods and Patterns: Innovating Information and Communication Technology*. Auerbach Publications, Boston, MA.

Weill, P. & Woerner, S. L. (2015). Thriving in an Increasingly Digital Ecosystem, *MIT Sloan Management Review*, 56 (4), 27-34.

Yang, H., Sloan, M. and Wang, J. 2015. Dynamic Information Retrieval Modeling, *WSDM'15, February 2–6, 2015, Shanghai, China. ACM* 978-1-4503-3317-7/15/02. http://dx.doi.org/10.1145/2684822.2697038.

Yue, Y. (2011). *New learning frameworks for information retrieval*, (PhD Thesis, Faculty of the Graduate School of Cornell University, NY, USA). Retrieved from http://www.yisongyue.com/yue_thesis.pdf

Zhu, D., Nimmagadda, S.L. & Reiners, T. (2018). An Integrated Information Retrieval Framework for Managing the Digital Web Ecosystem, Australasian Conference of Information Systems (ACIS, 2018), UTS, Sydney, Australia. http://www.acis2018.org/wp-content/uploads/2018/11/ACIS2018_paper_12.pdf

Zuccon, G., Leelanupab, T., Whiting, S., Yilmaz, E., Jose, J. M. & Azzopardi, L. (2013). Crowdsourcing interactions: using crowdsourcing for evaluating interactive information retrieval systems, *Information Retrieval Journal* (2013) 16:267–305. https://doi 10.1007/s10791-012-9206-z

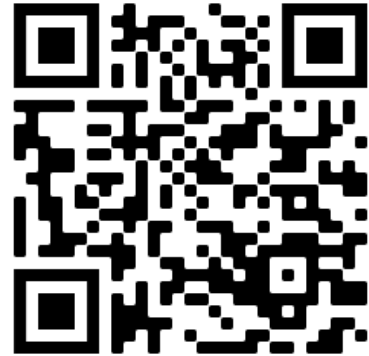